\documentclass[preprint,secnumarabic,showpacs,amsmath,amssymb,prd]{revtex4}%
\usepackage{amsfonts}
\usepackage{amsmath}
\usepackage{amssymb}
\usepackage{graphicx}%
\begin{document}
\title{Noncommutative 6D Gauge Higgs Unification Models}
\author{J. C. L\'{o}pez-Dom\'{\i}nguez}
\email{jlopez@fisica.ugto.mx}
\affiliation{Instituto de F\'{\i}sica de la Universidad de Guanajuato, P.O. Box E-143,
37150 Le\'on Gto., M\'exico.}
\author{O. Obreg\'on}
\email{octavio@fisica.ugto.mx}
\affiliation{Instituto de F\'{\i}sica de la Universidad de Guanajuato, P.O. Box E-143,
37150 Le\'on Gto., M\'exico.}
\author{C. Ram\'{\i}rez}
\email{cramirez@fcfm.buap.mx}
\affiliation{Facultad de Ciencias F\'{\i}sico Matem\'{a}ticas, Universidad Aut\'{o}noma de
Puebla, P.O. Box 1364, 72000 Puebla, M\'{e}xico.}
\author{J. J. Toscano}
\email{jtoscano@fcfm.buap.mx}
\affiliation{Facultad de Ciencias F\'{\i}sico Matem\'{a}ticas, Universidad Aut\'{o}noma de
Puebla, P.O. Box 1364, 72000 Puebla, M\'{e}xico.}

\date{\today}

\begin{abstract}
The influence of higher dimensions in noncommutative field theories
is considered. For this purpose, we analyze the bosonic sector of a
recently proposed 6 dimensional SU(3) orbifold model for the
electroweak interactions. The corresponding noncommutative theory is
constructed by means of the Seiberg-Witten map in 6D. We find in the
reduced bosonic interactions in 4D theory, couplings which are new
with respect to other known 4D noncommutative formulations of the
Standard Model using the Seiberg-Witten map. Phenomenological
implications due to the noncommutativity of extra dimensions are
explored. In particular, assuming that the commutative model leads
to the standard model values, a bound $-5.63\times 10^{-8}\
GeV^{-2}<\theta^{45}<1.06\times 10^{-7} \ GeV^{-2}$ on the
corresponding noncommutativity scale is derived from current
experimental constraints on the $S$ and $T$ oblique parameters. This
bound is used to predict a possibly significant impact of
noncommutativity effects of extra dimensions on the rare Higgs boson
decay $H\to\gamma\gamma$.

\end{abstract}
\pacs{11.10.Nx, 12.10.-g, 12.60.-i}
\maketitle

%%%%%%%%%%%%%%%%%%%%%%%%%%%%%%%%%%%%%%%%%%%%%%%%%%%%%%%%%%%%%%%%%%%%%%%%%%%

\section{Introduction}

A renewed interest in theories in 6D has recently emerged \cite{cgp}. An
anomaly free gauged supergravity in $D=6$, the Salam-Sezgin model \cite{ss},
has been considered. This model is compactified on a 2-sphere and in four
dimensions gives a $SU(2)\times U(1)$ gauge theory \cite{rss}. In particular,
it has been argued that these theories with 3-Branes could point out towards
solving the cosmological constant problem \cite{quevedo}. Also, in
\cite{brandenberger} it is shown that chaotic inflation consistent with
constraints coming from the amplitude of the cosmic microwave anisotropies can
be naturally realized.

In the search for a unified theory of elementary particles, the
incorporation of the Higgs field in the standard model (SM) of
electroweak interactions has motivated various proposals in 6D
\cite{Fairlie}. These are 6D pure gauge theories, in which after
dimensional reduction the Higgs field naturally arises. Recently new
proposals have been made, considering orbifold compactifications, in
\cite{antoniadis2}, a $U(3)\times U(3)$ model has been considered.
In this work the Higgs mass term is generated radiatively, with a
finite value at one loop as the quadratic divergences are suppressed
by the six dimensional gauge symmetry. Further, a $SU(3)$ model of
this type has been developed in \cite{scruca, Wulzer}, with one
Higgs doublet and a predicted W-boson mass of half the Higgs mass.
In this case the weak angle has a non realistic value, although it
can be improved by an extended gauge group as in \cite{antoniadis2},
or by the introduction of an $U(1)$ factor as done in \cite{scruca}.

Noncommutativity in field theories has been the subject of an
important number of works in the last few years. In particular, the
Seiberg-Witten construction \cite{SW} and its generalization for any
gauge group \cite{W2} have been studied. This construction allows to
express the noncommutative gauge fields in terms of the usual ones
and their derivatives, maintaining the same degrees of freedom. It
has been extended for noncommutative matter fields, which can also
be generated in terms of the commutative matter fields and the gauge
fields of interest \cite{W2}. By this procedure, noncommutative
versions of the standard model and consequently the electroweak
interaction sector have been given \cite{W3} (see also
\cite{Aschieri}). As a consequence new interactions among the fields
of the theory are predicted.

In this work, we will investigate the noncommutative generalization
of the bosonic sector of Gauge Higgs unification models in 6D based
on the $SU(3)$ gauge group compactified on $T^{2}/Z_{N}$
\cite{Wulzer}. The noncommutative extension is obtained by means of
the Seiberg-Witten map. We calculate, for the bosonic sector, the
resulting first order corrections and compare them with the results
obtained in other works. Assuming noncommutativity only between the
extra dimensions, the phenomenological consequences are considered
in the framework of the effective theories technique. First we
compute the new physics effects corrections to the $S$ and $T$
oblique parameters. Further, from the experimental constraints we
get a bound for the noncommutativity parameter $\theta^{45}$. This
bound allows us to calculate the correction to the decay width for
the rare decay of the Higgs boson into two photons.

In section 2 we review the model of reference \cite{Wulzer}, in section 3 the
Seiberg-Witten map and its generalization to nonabelian groups is presented in
some detail. In section 4 we present the noncommutative formulation of the
model \cite{Wulzer} and show that our results differ from those calculated
directly in 4D. The phenomenological consequences of the noncommutativity
between extra dimensions are discussed in section 5. Section 6 is devoted to conclusions.

%%%%%%%%%%%%%%%%%%%%%%%%%%%%%%%%%%%%%%%%%%%%%%%%%%%%%%%%%%%%%%%%%%%%%%%%%%%%%%

\section{\noindent The 6-Dimensional Model}

\subsection{Gauge Fields in a 6-Dimensional Space-Time}

Let us consider a Yang-Mills theory in 6-dimensional space-time with a $SU(3)$
gauge group, the Lagrangian of the theory is
\begin{equation}
\mathcal{L}=-\frac{1}{2}\mathrm{Tr}F_{mn}F^{mn},\nonumber
\end{equation}
the field strength tensor is defined by%
\begin{equation}
F_{mn}=\partial_{m}A_{n}-\partial_{n}A_{m}-ig_{6}[A_{m},A_{n}],\nonumber
\end{equation}
and $g_{6}$ is the coupling constant in 6D. This action is interpreted by a
dimensional reduction on an orbifold $T^{2}/Z_{N}$ for $N=3,4,6$
\cite{Wulzer}, by a separation of the connection in its 4-dimensional
space-time part $A_{\mu}$, and the other two components, $A_{z}$ and
$A_{\overline{z}}$ which in 4D will play the role of scalars, with $z=\frac
{1}{\sqrt{2}}\left(  x^{4}+ix^{5}\right)  $ and $\overline{z}=\frac{1}%
{\sqrt{2}}\left(  x^{4}-ix^{5}\right)  $. These fields are the zero modes of
Kaluza-Klein and depend only on the four space-time coordinates $x^{\mu}$. The
result of this reduction is given by%

\begin{equation}
\mathcal{L}=-\frac{1}{2}\mathrm{Tr}F_{\mu\nu}F^{\mu\nu}+2\mathrm{Tr}D_{\mu
}A_{\overline{z}}D^{\mu}A_{z}-g^{2}\mathrm{Tr}\left[  A_{z},A_{\overline{z}%
}\right]  ^{2}, \label{lagc}%
\end{equation}
where $g=g_{6}\sqrt{V}$ is the gauge coupling of the 4-dimensional effective
theory, $V$ is the volume of the two extra dimensions and%
\begin{align}
F_{\mu\nu}  &  =\partial_{\mu}A_{\nu}-\partial_{\nu}A_{\mu}-ig[A_{\mu},A_{\nu
}],\nonumber\\
D_{\mu}A_{z}  &  =\partial_{\mu}A_{z}-ig[A_{\mu},A_{z}]=F_{\mu z},\\
D_{\mu}A_{\overline{z}}  &  =\partial_{\mu}A_{\overline{z}}-ig[A_{\mu
},A_{\overline{z}}]=F_{\mu\overline{z}}\,.\nonumber
\end{align}
The orbifold reduction \cite{Wulzer} for the gauge fields $A_{m}$ leads to:
the 4-dimensional $A_{\mu},$ that containts four electroweak bosons, $W_{\mu
}\in SU\left(  2\right)  $ and $B_{\mu}\in U\left(  1\right)  $,
\[
A_{\mu}=\left(
\begin{array}
[c]{cc}%
W_{\mu} & 0\\
0 & 0
\end{array}
\right)  +\frac{1}{2\sqrt{3}}\left(
\begin{array}
[c]{cc}%
B_{\mu}I & 0\\
0 & -2B_{\mu}%
\end{array}
\right)  ,
\]
and the two complex components of the scalar boson doublet (Higgs), which are
contained in the $A_{z}$ and $A_{\overline{z}}$ gauge fields,
\[
A_{z}=\frac{1}{\sqrt{2}}\left(
\begin{tabular}
[c]{ll}%
$0$ & $\phi$\\
$0$ & $0$%
\end{tabular}
\right)  ,\;\;\;A_{\overline{z}}=\frac{1}{\sqrt{2}}\left(
\begin{tabular}
[c]{ll}%
$0$ & $0$\\
$\phi^{\dag}$ & $0$%
\end{tabular}
\right)  .
\]

\noindent Substituting these expressions in the Lagrangian (\ref{lagc}) we find%
\begin{equation}
\mathcal{L}=-\frac{1}{2}\mathrm{Tr}F_{\mu\nu}(W)F^{\mu\nu}(W)-\frac{1}%
{4}F_{\mu\nu}(B)F^{\mu\nu}(B)+\left(  D_{\mu}\phi\right)  ^{\dag}(D^{\mu}%
\phi)-V\left(  \phi\right)  , \label{lagrangiano1}%
\end{equation}
where $D_{\mu}\phi=\left(  \partial_{\mu}-\frac{1}{2}igW_{\mu}^{a}\tau
_{a}-\frac{1}{2}ig\tan\theta_{W}B_{\mu}\right)  \phi,$ $\tan\theta_{W}%
=\sqrt{3}$ and $V\left(  \phi\right)  =\frac{g^{2}}{2}\left\vert
\phi\right\vert ^{4}.$

\noindent Thus, this Lagrangian has a $SU(2)\times U(1)$ invariance, with a
scalar massless doublet with a quartic potential. However, as shown in
\cite{antoniadis1}, quantum fluctuations induce corrections to the potential
$V\left(  \phi\right)  $ which can trigger radiative symmetry breaking. The
leading terms in the one-loop effective potential for the Higgs are,
\[
V_{eff}\left(  \phi\right)  =-\mu^{2}\left\vert \phi\right\vert ^{2}%
+\lambda\left\vert \phi\right\vert ^{4}.
\]

\noindent Assuming $\mu^{2}>0,$ so that electroweak symmetry breaking can
occur, we have that $\left\langle \left\vert \phi\right\vert \right\rangle
=\nu/\sqrt{2}$ with $\nu=\mu/\sqrt{\lambda}$. Note that the value of the
electroweak angle in (\ref{lagrangiano1}) is too large. However, as mentioned
in the introduction, the model can be extended in such a way that it correctly
reproduces the SM value.

%%%%%%%%%%%%%%%%%%%%%%%%%%%%%%%%%%%%%%%%%%%%%%%%%%%%%%%%%%%%%%%%%%%%%%%%%%%%%%

\section{\noindent Noncommutative Gauge Theories}

\subsection{Noncommutative Space-Time}

Noncommutative space-time incorporates coordinates $\widehat{x}^{\mu}$, given
by operators that satisfy the following relations,
\begin{equation}
\left[  \widehat{x}^{\mu},\widehat{x}^{\nu}\right]  =i\theta^{\mu\nu},
\label{canonicalNC}%
\end{equation}
where $\theta^{\mu\nu}=-\theta^{\nu\mu}$ are real numbers. To construct a
field theory in this space, it is more convenient to consider usual fields,
which are functions. This is allowed by the Weyl-Wigner-Moyal correspondence,
which establishes an equivalence between the Heisenberg algebra of the
operators $\widehat{x}^{\mu}$ and the function algebra in $\mathbb{R}^{m}$. It
has an associative and noncommutative star product, the Moyal $\star$-product,
given by,
\begin{align}
f(x)\star g(x)  &  \equiv\left[  \exp\left(  \frac{i}{2}\frac{\partial
}{\partial\varepsilon^{\alpha}}\theta^{\alpha\beta}\frac{\partial}%
{\partial\eta^{\beta}}\right)  f(x+\varepsilon)g(y+\eta)\right]
_{\varepsilon=\eta=0}\label{starproduct}\\
&  =fg+\frac{i}{2}\theta^{\alpha\beta}\,\partial_{\alpha}f\,\partial_{\beta
}g+\mathcal{O}(\theta^{2}).\nonumber
\end{align}
Under complex conjugation it satisfies $\left(  \overline{f\star
g}\right) =\overline{g}\star\overline{f}$. Since we will work with a
nonabelian gauge group, our functions are matrix valued, and the
corresponding matrix Moyal product is denoted by an $\ast$. In this
case the hermitian conjugation is given by $\left(  f\ast g\right)
^{\dagger}=g^{\dagger}\ast f^{\dagger}$. Under the integral, for
closed manifolds, this product has the cyclic property
$\mathrm{Tr}\int f_{1}\ast f_{2}\ast\cdot\cdot\cdot\ast
f_{n}=\mathrm{Tr}\int f_{n}\ast f_{1}\ast
f_{2}\ast\cdot\cdot\cdot\ast f_{n-1}$. In particular
$\mathrm{Tr}\int f_{1}\ast f_{2}=\mathrm{Tr}\int f_{1}f_{2}.$
Therefore, a theory on the noncommutative space of the
$\widehat{x}$, is equivalent to a theory of usual fields, where the
function product is substituted by the Moyal $\ast$-product.

This suggests that any theory can be converted into a noncommutative one by
replacing the ordinary function product with the $\ast$-product.

%%%%%%%%%%%%%%%%%%%%%%%%%%%%%%%%%%%%%%%%%%%%%%%%%%%%%%%%%%%%%%%%%%%%%%%%%%%%%%%%%%%%

\subsection{The Seiberg-Witten Map}

In order to build noncommutative Yang-Mills theories it is necessary, first of
all, that a commutative limit exists and that a perturbative study of the
noncommutative theory is possible. In this case the solutions of such a theory
must depend on the noncommutativity parameter $\theta$ in the form of a power
series expansion.

For an ordinary Yang-Mills theory, the gauge field and the strength field
tensor transformations can be written as:
\begin{align}
\delta_{\lambda}A_{\mu}  &  =\partial_{\mu}\lambda+i\lambda A_{\mu}-iA_{\mu
}\lambda,\label{deltaAmu}\\
F_{\mu\nu}  &  =\partial_{\mu}A_{\nu}-\partial_{\nu}A_{\mu}-iA_{\mu}A_{\nu
}+iA_{\nu}A_{\mu},\nonumber\\
\delta_{\lambda}F_{\mu\nu}  &  =i\lambda F_{\mu\nu}-iF_{\mu\nu}\lambda
.\nonumber
\end{align}
For the noncommutative gauge theory, we use the same equations in the gauge
field and strength field tensor transformations, except that the matrix
multiplications are replaced by the $\ast$ product. Then the gauge field and
the strength field tensor transformations are \cite{SW}:
\begin{align}
\widehat{\delta}_{\hat{\lambda}}\widehat{A}_{\mu}  &  =\partial_{\mu}%
\widehat{\lambda}+i\widehat{\lambda}\ast\widehat{A}_{\mu}-i\widehat{A}_{\mu
}\ast\widehat{\lambda},\nonumber\\
\widehat{F}_{\mu\nu}  &  =\partial_{\mu}\widehat{A}_{\nu}-\partial_{\nu
}\widehat{A}_{\mu}-i\widehat{A}_{\mu}\ast\widehat{A}_{\nu}+i\widehat{A}_{\nu
}\ast\widehat{A}_{\mu},\label{transAnc}\\
\widehat{\delta}_{\hat{\lambda}}\widehat{F}_{\mu\nu}  &  =i\widehat{\lambda
}\ast\widehat{F}_{\mu\nu}-i\widehat{F}_{\mu\nu}\ast\widehat{\lambda},\nonumber
\end{align}
from which the original Yang-Mills theory (\ref{deltaAmu}) results in the
limit $\theta\rightarrow0$. Notice that equations (\ref{transAnc}) are valid
even for abelian gauge fields. Due to the form of the Moyal product
(\ref{starproduct}), the noncommutative theory has the structure of a nonlocal theory.
However, if we consider it as an effective theory, its energy scale gives us a
cutoff and nonlocality is not a problem.
Further, as shown by Kontsevich \cite{Kontsevich}, at the level of the
physical degrees of freedom there is a one to one relation between the
commutative and the noncommutative theories. However both theories
are quite different, as noncommutativity generates new couplings.

Let us consider the noncommutative gauge transformations of an abelian theory,
\begin{equation}
\delta_{\widehat{{\lambda}}}\widehat{A}_{\mu}=\partial_{\mu}\widehat{\lambda
}+i\widehat{\lambda}\star\widehat{A}_{\mu}-i\widehat{A}_{\mu}\star
\widehat{\lambda},
\end{equation}
we see that they look like nonabelian ones, although they continue to depend
on only one generator. For nonabelian groups, things are more complicated
\cite{W2},
\begin{align}
\delta_{\widehat{{\lambda}}}\widehat{A}_{\mu}  &  =\partial_{\mu}%
\widehat{\lambda}+i\widehat{\lambda}\ast\widehat{A}_{\mu}-i\widehat{A}_{\mu
}\ast\widehat{\lambda}\nonumber\\
&  =\partial_{\mu}\widehat{\lambda}^{a}\Lambda_{a}+i\widehat{\lambda}%
^{a}\Lambda_{a}\ast\widehat{A}_{\mu}^{b}\Lambda_{b}-i\widehat{A}_{\mu}%
^{b}\Lambda_{b}\ast\widehat{\lambda}^{a}\Lambda_{a}\nonumber\\
&  =\partial_{\mu}\widehat{\lambda}^{a}\Lambda_{a}+\frac{i}{2}\left\{
\widehat{\lambda}^{a}\overset{\star}{,}\widehat{A}_{\mu}^{b}\right\}  \left[
\Lambda_{a},\Lambda_{b}\right]  +\frac{i}{2}\left[  \widehat{\lambda}%
^{a}\overset{\star}{,}\widehat{A}_{\mu}^{b}\right]  \left\{  \Lambda
_{a},\Lambda_{b}\right\}  . \label{con y anticon}%
\end{align}
Now the transformation algebra is generated by commutators and
anticommutators, which amounts to the universal enveloping algebra of the
original algebra $U(g,R)$, where $R$ is the corresponding representation. The
generators of this algebra satisfy,
\begin{equation}
\lbrack\Lambda_{A},\Lambda_{B}]=if_{ABC}\Lambda_{C}\ ,\qquad\{\Lambda
_{A},\Lambda_{B}\}=d_{ABC}\Lambda_{C}, \label{amo10}%
\end{equation}
where $f_{ABC}=-f_{BAC}$ and $d_{ABC}=d_{BAC}$ are the structure constants.

\noindent These transformations are satisfied order by order on $\theta$, and
all coefficients of the higher terms can be used to fix the gauge degrees of
freedom of $\widehat{A}_{\mu}$. In such a gauge fixing, the only remaining
freedom of the transformation parameters $\widehat{\lambda}$ are the ones of
the commutative theory, so they should depend only on $\lambda^{a}$ and their
derivatives. In this case consistency implies that an infinitesimal
commutative gauge transformation $\delta_{\lambda}{A}_{\mu}=\partial_{\mu
}{\lambda}+i{\lambda}{A}_{\mu}-i{A}_{\mu}{\lambda}$, will induce the
noncommutative one,
\begin{equation}
\widehat{A}_{\mu}(A+\delta_{\lambda}A)=\widehat{A}_{\mu}(A)+\widehat{\delta
}_{\hat{\lambda}}\widehat{A}_{\mu}(A). \label{sw1}%
\end{equation}
This is the so called Seiberg-Witten map.

The solution to (\ref{sw1}) can be obtained by setting $\widehat{A}_{\mu
}=A_{\mu}+A_{\mu}^{\prime}(A)$ and $\widehat{\lambda}=\lambda+\lambda^{\prime
}(\lambda,A)$, where $A_{\mu}^{\prime}$ and $\lambda^{\prime}$ are local
functions of $\lambda$ and $A_{\mu}$ of first order in $\theta$. Then
substituting in (\ref{sw1}) and expanding to first order,
\begin{equation}
A_{\mu}^{\prime}(A+\delta_{\lambda}A)-A_{\mu}^{\prime}(A)-\partial_{\mu
}\lambda^{\prime}-i[\lambda^{\prime},A_{\mu}]-i[\lambda,A_{\mu}^{\prime
}]=-\frac{1}{2}\theta^{\alpha\beta}(\partial_{\alpha}\lambda\partial_{\beta
}A_{\mu}+\partial_{\beta}A_{\mu}\partial_{\alpha}\lambda). \label{sw2}%
\end{equation}
One solution of this equation is given by \cite{SW},
\begin{align}
\widehat{A}_{\mu}(A)  &  =A_{\mu}+A_{\mu}^{\prime}(A)=A_{\mu}-\frac{1}%
{4}\theta^{\alpha\beta}\left\{  A_{\alpha},\partial_{\beta}A_{\mu}+F_{\beta
\mu}\right\}  +\mathcal{O}(\theta^{2}),\label{Anc}\\
\widehat{\lambda}(\lambda,A)  &  =\lambda+\lambda^{\prime}(\lambda
,A)=\lambda+\frac{1}{4}\theta^{\alpha\beta}\left\{  \partial_{\alpha}%
\lambda,A_{\beta}\right\}  +\mathcal{O}(\theta^{2}), \label{lambdanc}%
\end{align}
from which it turns out that,
\begin{equation}
\widehat{F}_{\mu\nu}=F_{\mu\nu}+\frac{1}{4}\theta^{\alpha\beta}\left(
2\left\{  F_{\mu\alpha},F_{\nu\beta}\right\}  -\left\{  A_{\alpha},\left(
D_{\beta}+\partial_{\beta}\right)  F_{\mu\nu}\right\}  \right)  +\mathcal{O}%
(\theta^{2}). \label{Fnc2}%
\end{equation}

\noindent These equations (\ref{Anc}, \ref{lambdanc}, \ref{Fnc2}) are the
explicit form of the Seiberg-Witten map, which in this way can be constructed
for any Lie algebra of transformations \cite{W2}.

As shown the noncommutative generators $\widehat\lambda$ take values in the
enveloping algebra. In the case of the fundamental representation of unitary
groups $U(N)$, they coincide with their enveloping algebras. For the algebra
of $SU(N)$ in the fundamental representation, the enveloping algebra
incorporates, through the anticommutators of the generators, the identity
matrix $\Lambda_{0}=\frac{1}{\sqrt{2N}}\mathbb{I}_{N\times N}$, and is then
given by $U(N)$.

%%%%%%%%%%%%%%%%%%%%%%%%%%%%%%%%%%%%%%%%%%%%%%%%%%%%%%%%%%%%%%%%%%%%%%%%%%%%%%%%%

\section{\noindent The Noncommutative Model}

As previously mentioned, our purpose is the construction of a noncommutative
version of the 6-dimensional $SU(3)$ gauge theory presented in Section 2. The
fact that we are considering noncommutativity in 6D, means that we only need
the Seiberg-Witten map for gauge fields. Thus the effects of noncommutativity
on the Higgs field and its interactions will arise after dimensional
reduction, in particular from the Seiberg-Witten map of the gauge fields
$A_{z}$ and $A_{\overline{z}}$.

The noncommutative action is given by:
\begin{equation}
\widehat{\mathcal{S}}_{NC}=-\frac{1}{2}\mathrm{Tr}\int d^{6}x\widehat{F}%
_{mn}\widehat{F}^{mn}, \label{NCaction}%
\end{equation}
where%
\begin{equation}
\widehat{F}_{mn}=F_{mn}+\frac{1}{4}\theta^{kl}\left(  2\left\{  F_{mk}%
,F_{nl}\right\}  -\left\{  A_{k},\left(  D_{l}+\partial_{l}\right)
F_{mn}\right\}  \right)  +\mathcal{O}(\theta^{2}). \label{Fnc3}%
\end{equation}
Here the indexes $m,n,k$ and $l$ take the values $0,...,3,z$ and $\overline
{z}$. Thus the noncommutative parameter $\theta^{kl}$ can be: $\theta^{\mu\nu
}$ (noncommutativity among the 4-dimensional space-time coordinates),
$\theta^{\mu z},$ $\theta^{\mu\overline{z}}$ (noncommutativity among the
4-dimensional space-time coordinates and the extra dimensions coordinates) and
$\theta^{z\overline{z}}$ (noncommutativity between the extra dimensions).
Therefore, after inserting the noncommutative field strength (\ref{Fnc3}) into
(\ref{NCaction}), the noncommutative action gets the following first order
corrections,
\begin{align}
&  -\frac{\theta^{\alpha\beta}}{4}\mathrm{Tr}\Bigg\{\Big[2\left\{
F_{\mu\alpha},F_{\nu\beta}\right\}  -\left\{  A_{\alpha},\left(  D_{\beta
}+\partial_{\beta}\right)  F_{\mu\nu}\right\}  \Big]F^{\mu\nu}\nonumber\\
&  \qquad\qquad+2\Big[2\left\{  F_{\mu\alpha},F_{z\beta}\right\}  -\left\{
A_{\alpha},\left(  D_{\beta}+\partial_{\beta}\right)  F_{\mu z}\right\}
\Big]F^{\mu z}\nonumber\\
&  \qquad\qquad+2\Big[2\left\{  F_{\mu\alpha},F_{\overline{z}\beta}\right\}
-\left\{  A_{\alpha},\left(  D_{\beta}+\partial_{\beta}\right)  F_{\mu
\overline{z}}\right\}  \Big]F^{\mu\overline{z}}\nonumber\\
&  \qquad\qquad+2\Big[2\left\{  F_{z\alpha},F_{\overline{z}\beta}\right\}
-\left\{  A_{\alpha},\left(  D_{\beta}+\partial_{\beta}\right)  F_{z\overline
{z}}\right\}  \Big]F^{z\overline{z}}\Bigg\}\nonumber\\
&  -\frac{\theta^{\alpha i}}{4}\mathrm{Tr}\Bigg\{\Big[4\left\{  F_{\mu\alpha
},F_{\nu i}\right\}  -\left\{  A_{\alpha},\left(  D_{i}+\partial_{i}\right)
F_{\mu\nu}\right\}  +\left\{  A_{i},\left(  D_{\alpha}+\partial_{\alpha
}\right)  F_{\mu\nu}\right\}  \Big]F^{\mu\nu}\nonumber\\
&  \qquad\qquad+2\Big[2\{F_{\mu\alpha},F_{ji}\}-2\{F_{j\alpha},F_{\mu
i}\}-\left\{  A_{\alpha},\left(  D_{i}+\partial_{i}\right)  F_{\mu j}\right\}
+\left\{  A_{i},\left(  D_{\alpha}+\partial_{\alpha}\right)  F_{\mu
j}\right\}  \Big]F^{\mu j}\nonumber\\
&  \qquad\qquad+2\Big[2\{F_{i\alpha},F_{\overline{z}z}\}-\left\{  A_{\alpha
},\left(  D_{i}+\partial_{i}\right)  F_{z\overline{z}}\right\}  +\left\{
A_{i},\left(  D_{\alpha}+\partial_{\alpha}\right)  F_{z\overline{z}}\right\}
\Big]F^{z\overline{z}}\Bigg\}\nonumber\\
&  -\frac{\theta^{z\overline{z}}}{4}\mathrm{Tr}\Bigg\{\Big[4\left\{  F_{\mu
z},F_{\nu\overline{z}}\right\}  -\left\{  A_{z},D_{\overline{z}}F_{\mu\nu
}\right\}  +\left\{  A_{\overline{z}},D_{z}F_{\mu\nu}\right\}  \Big]F^{\mu\nu
}\nonumber\\
&  \qquad\qquad+2\Big[2\{F_{\mu z},F_{z\overline{z}}\}-\left\{  A_{z}%
,D_{\overline{z}}F_{\mu z}\right\}  +\left\{  A_{\overline{z}},D_{z}F_{\mu
z}\right\}  \Big]F^{\mu z}\nonumber\\
&  \qquad\qquad+2\Big[2\{F_{z\overline{z}},F_{\mu\overline{z}}\}-\left\{
A_{z},D_{\overline{z}}F_{\mu\overline{z}}\right\}  +\left\{  A_{\overline{z}%
},D_{z}F_{\mu\overline{z}}\right\}  \Big]F^{\mu\overline{z}}\nonumber\\
&  \qquad\qquad+2\Big[2\{F_{z\overline{z}},F_{z\overline{z}}\}-\left\{
A_{z},D_{\overline{z}}F_{z\overline{z}}\right\}  +\left\{  A_{\overline{z}%
},D_{z}F_{z\overline{z}}\right\}  \Big]F^{z\overline{z}}\Bigg\},
\label{lagrangianoNC1}%
\end{align}
where $\mu,\nu,\alpha,\beta=0,...,3$, $i=z,\overline{z}$.

\noindent After somewhat cumbersome computations, we obtain the following
expression for these corrections in terms of the $SU(2)$ and $U(1)$ field
strengths $W^{\mu\nu}$ and $B^{\mu\nu}$ respectively, the corresponding gauge
fields $W^{\mu}$ and $B^{\mu}$ and the Higgs field $\phi$,
\begin{align}
\widehat{\mathcal{L}}_{NC}  &  =-\frac{1}{2}\mathrm{Tr}\,W_{\mu\nu}W^{\mu\nu
}-\frac{1}{4}B_{\mu\nu}B^{\mu\nu}+\left(  D_{\mu}\phi\right)  ^{\dagger
}(D^{\mu}\phi)-\frac{g^{2}}{2}\left\vert \phi\right\vert ^{4}\nonumber\\
&  -\frac{1}{4}\theta^{\alpha\beta}\Bigg\{\frac{1}{2\sqrt{3}}\mathrm{Tr}%
\Big[4\{W_{\mu\alpha},B_{\nu\beta}\}W^{\mu\nu}+2\{W_{\mu\alpha},W_{\nu\beta
}\}B^{\mu\nu}I-\{W_{\alpha},D_{\beta}W_{\mu\nu}\}B^{\mu\nu}I\nonumber\\
&  \qquad\qquad-\{B_{\alpha},D_{\beta}W_{\mu\nu}\}W^{\mu\nu}\Big]+\frac
{1}{2\sqrt{3}}B_{\alpha}\partial_{\beta}B_{\mu\nu}B^{\mu\nu}-\frac{1}%
{2\sqrt{3}}B_{\mu\alpha}B_{\nu\beta}B^{\mu\nu}\nonumber\\
&  \qquad\qquad+2(D^{\mu}\phi)^{\dagger}\Big(W_{\mu\alpha}-\frac{1}{2\sqrt{3}%
}B_{\mu\alpha}I\Big)(D_{\beta}\phi)+h.c.\nonumber\\
&  \qquad\qquad+(D^{\mu}\phi)^{\dagger}\Big(W_{\alpha}-\frac{1}{2\sqrt{3}%
}B_{\alpha}I\Big)\Big(\overrightarrow{\partial}_{\beta}+\overrightarrow
{D}_{\beta}\Big)(D_{\mu}\phi)\nonumber\\
&  \qquad\qquad+(D^{\mu}\phi)^{\dagger}\Big(\overleftarrow{\partial}_{\beta
}+\overleftarrow{D}_{\beta}\Big)\Big(W_{\alpha}-\frac{1}{2\sqrt{3}}B_{\alpha
}I\Big)(D_{\mu}\phi)\nonumber\\
&  \qquad\qquad+ig\Big[\phi^{\dagger}(D_{\alpha}\phi)(D_{\beta}\phi)^{\dagger
}\phi-(D_{\beta}\phi)^{\dagger}(D_{\alpha}\phi)\phi^{\dagger}\phi
\Big]-ig^{3}\phi^{\dagger}\phi\phi^{\dagger}W_{\beta}W_{\alpha}\phi\nonumber\\
&  \qquad\qquad-g^{2}\Big[\phi^{\dagger}\Big(W_{\alpha}+\frac{1}{2\sqrt{3}%
}B_{\alpha}I\Big)\partial_{\beta}(\phi\phi^{\dagger})\phi-\frac{2}{\sqrt{3}%
}B_{\alpha}\partial_{\beta}(\phi\phi^{\dagger})\phi^{\dagger}\phi\nonumber\\
&  \qquad\qquad+\phi^{\dagger}\partial_{\beta}(\phi\phi^{\dagger
})\Big(W_{\alpha}+\frac{1}{2\sqrt{3}}B_{\alpha}I\Big)\phi
\Big]\Bigg\}\nonumber\\
&  +\frac{i}{2}\theta^{z\overline{z}}\Bigg\{-2i(D_{\mu}\phi)^{\dagger
}\Big(W^{\mu\nu}+\frac{1}{\sqrt{3}}B^{\mu\nu}I\Big)(D_{\nu}\phi)\nonumber\\
&  \qquad\qquad+\frac{g}{2}\Big[\phi^{\dagger}\phi(D_{\mu}\phi)^{\dagger
}(D^{\mu}\phi)-(D_{\mu}\phi)^{\dagger}\phi\phi^{\dagger}(D^{\mu}%
\phi)\Big]\nonumber\\
&  \qquad\qquad-g\,\phi^{\dagger}\Big(W_{\mu\nu}W^{\mu\nu}+\frac{1}{\sqrt{3}%
}W_{\mu\nu}B^{\mu\nu}-\frac{1}{4}B_{\mu\nu}B^{\mu\nu}\Big)\phi\Bigg\},
\label{lagnc}%
\end{align}

\noindent In this equation there are new interactions with respect
to the ones found in the $4D$ noncommutative formulations of the
SM \cite{W3,Chaichian1}. For instance the interactions between the
weak gauge fields and the electromagnetic field which appear in
the first terms that multiply the four dimensional
noncommutativity parameter $\theta^{\alpha\beta }$. Note that
there are not corrections linear in the noncommutativity parameter
$\theta^{i\alpha}$, they turn out to be identically zero, as a
consequence of the orbifold symmetries. Of particular interest are
the corrections corresponding to noncommutativity between the
extra dimensions, i.e. the terms multiplied by
$\theta^{z\overline{z}}$, given by interactions among the Higgs
and the gauge bosons, and also higher order Higgs
self-interactions. Considering only these sort of corrections, we
have,
\begin{align}
\widehat{\mathcal{L}}_{NC}=  &  -\frac{1}{2}\mathrm{Tr}\,W_{\mu\nu}W^{\mu\nu
}-\frac{1}{4}B_{\mu\nu}B^{\mu\nu}+\left(  D_{\mu}\phi\right)  ^{\dagger
}(D^{\mu}\phi)-\frac{g^{2}}{2}\left\vert \phi\right\vert ^{4}\nonumber\\
&  +\frac{i}{2}\theta^{z\overline{z}}\Bigg\{-2i\left(  D_{\mu}\phi\right)
^{\dagger}\Big(W^{\mu\nu}+\frac{1}{\sqrt{3}}B^{\mu\nu}I\Big)(D_{\nu}%
\phi)\nonumber\\
&  \qquad\qquad+\frac{g}{2}\big[\phi^{\dagger}\phi\left(  D_{\mu}\phi\right)
^{\dagger}(D^{\mu}\phi)-\left(  D_{\mu}\phi\right)  ^{\dagger}\phi
\phi^{\dagger}(D^{\mu}\phi)\big]\nonumber\\
&  \qquad\qquad-g\,\phi^{\dagger}\Big(W_{\mu\nu}W^{\mu\nu}+\frac{1\,}{\sqrt
{3}}W_{\mu\nu}B^{\mu\nu}-\frac{1}{4}B_{\mu\nu}B^{\mu\nu}\Big)\phi\Bigg\}.
\label{lagnc1}%
\end{align}
The noncommutative corrections in this Lagrangian are dimension-six
operators, well known from the electroweak effective Lagrangian technique
\cite{EL}, a scheme in which the effects of these terms can be studied in a
model-independent manner.
It is interesting to consider the extension of the $SU(3)$ six dimensional
gauge group by an $U(1)$ factor. Its gauge field is invariant under the orbifold
symmetries \cite{scruca} and as it does not mix with the $SU(3)$ gauge field, the
noncommutative corrections (\ref{Fnc3}) and in (\ref{lagnc1}) will not be affected.

%%%%%%%%%%%%%%%%%%%%%%%%%%%%%%%%%%%%%%%%%%%%%%%%%%%%%%%%%%%%%%%%%%%%%%%%%%%%%%

\section{Phenomenological implications}

In this section we will analyze the phenomenological implications of
noncommutativity between extra dimensions, in the case of Lagrangian
(\ref{lagnc1}). Thus, we fix our attention on the terms multiplied
by $\theta^{45}$ $\left(  \theta^{45}=i\theta^{z\overline{z}}\right)
$. These terms contain new interactions relative to the SM and its
noncommutative versions \cite{W3, Chaichian1}. As mentioned, the
model we are considering predicts a too large weak angle. However we
can expect that the kind of noncommutative terms considered here
will also arise from a more realistic theory, that would reproduce
the SM in the limit of vanishing $\theta$ parameters. For instance,
as mentioned end of the last section, in the case of an $U(1)$
extension, the noncommutative corrections which affect the
electroweak gauge fields are not modified. Accordingly, in the
following we will consider these terms as representing deviations of
the genuine standard electroweak Lagrangian. In order to analyze
their effects, we observe that these new interactions are given by
well--known dimension--six operators, which have been already
studied in, by example, the electroweak effective Lagrangian
approach \cite{EL}, a scheme appropriate to investigate in a
model--independent manner physics lying beyond the Fermi scale. For
this purpose, it is convenient to divide the above operators into
three sets, as follows
\begin{align}
&  \mathcal{O}_{\phi W}=\frac{g}{2}\theta^{45}(\phi^{\dag}W_{\mu\nu}W^{\mu\nu
}\phi),\label{o1}\\
&  \mathcal{O}_{\phi B}=-\frac{g}{8}\theta^{45}(\phi^{\dag}B_{\mu\nu}B^{\mu
\nu}\phi),\label{o2}\\
&  \mathcal{O}_{WB}=\frac{g}{2\sqrt{3}}\theta^{45}(\phi^{\dag}W_{\mu\nu}%
B^{\mu\nu}\phi) ,\label{o3}%
\end{align}%
\begin{align}
&  \mathcal{O}_{\phi}^{(1)}=-g\theta^{45}(\phi^{\dag}\phi)(D_{\mu}\phi)^{\dag
}(D^{\mu}\phi),\\
&  \mathcal{O}_{\phi}^{(3)}=g\theta^{45}[(D_{\mu}\phi)^{\dag}\phi][\phi^{\dag
}(D^{\mu}\phi)],
\end{align}%
\begin{align}
&  \mathcal{O}_{DW}=i\theta^{45}(D_{\mu}\phi)^{\dag}W^{\mu\nu}(D_{\nu}\phi),\\
&
\mathcal{O}_{DB}=\frac{i}{\sqrt{3}}\theta^{45}(D_{\mu}\phi)^{\dag}B^{\mu
\nu}(D_{\nu}\phi).
\end{align}
First we observe that there are
potential modifications induced by these operators on the
quadratic SM Lagrangian, as they can alter some tree level
relations which are experimentally constrained, such as the
kinetic energy part of the $W$ and $Z$ bosons. In particular, they
can give tree--level contributions to the $S$ and $T$ oblique
parameters\cite{PT}. Let us focus our attention on those
interactions which affect the quadratic part of the SM gauge
sector.

After spontaneous symmetry breaking, all the above operators
induce new nonrenormalizable interactions, as well as
renormalizable ones, which modify those predicted by the
dimension--four theory. In particular, the first two sets of
operators induce bilinear terms that can eventually modify the SM
parameters \cite{EL,Wudka2}. On the other hand, although the last
set of operators are potentially interesting from the
phenomenological point of view \cite{TH}, they are not important
for our purposes, as they do not introduce modifications in the
quadratic Lagrangian. Concerning the first set, it is easy to see
that $\mathcal{O}_{\phi W}$ and $\mathcal{O}_{\phi B}$ modify the
canonical form of the kinetic terms $W_{\mu\nu}W^{\mu\nu}$ and
$B_{\mu\nu }B^{\mu\nu}$, respectively. However, these effects are
unobservable indeed, since they can be absorbed in a finite
renormalization of the gauge fields and the coupling constant $g$.
As to the $\mathcal{O}_{WB}$, $\mathcal{O}_{\phi }^{(1)}$, and
$\mathcal{O}_{\phi}^{(3)}$ operators, they introduce nontrivial
modifications in the quadratic Lagrangian. In particular, as we
will see below, the first and the last of these operators are
sensitive to the low--energy data, as they contribute to the $S$
and $T$ parameters at the tree level. Up to some surface terms,
the quadratic part of the effective Lagrangian, \textit{i.e.}, the
SM and new contributions, can conveniently be written as
\begin{eqnarray}
\label{lwb}
\mathcal{L}_{Kinetic}&=&\frac{1}{2}W^{a\mu}\Big\{\Big[\Box+
\frac{g^2v^2}{4}\Big(1-\frac{\alpha^{(1)}_\phi}{2}\Big)\Big]g_{\mu
\nu}-\partial_\mu \partial_\nu\Big\}W^{a\nu} \nonumber \\
&&+\frac{1}{2}B^{\mu}\Big\{\Big[\Box+
\frac{g'^2v^2}{4}\Big(1-\frac{\alpha^{(1)}_{\phi}}{2}\Big)\Big]g_{\mu
\nu}-\partial_\mu \partial_\nu\Big\}B^{\nu} \nonumber \\
&&+W^{3\mu}\Big(\frac{g^2v^2}{16}\alpha^{(3)}_\phi\Big)g_{\mu
\nu}W^{3\nu}+W^{3\mu}\Big[\alpha_{WB}\Big(\Box g_{\mu
\nu}-\partial_\mu \partial_\nu\Big)\Big]B^\nu,
\end{eqnarray}
where the unobservable effects arising from the $\mathcal{O}_{\phi
W}$ and $\mathcal{O}_{\phi B}$ operators were ignored. In
addition, in order to identify the origin of each contribution, we
have introduced the definitions:
$\alpha_{WB}=gv^2\theta^{45}/2\sqrt{3}$ and $\alpha^{(1)}_\phi=
\alpha^{(3)}_\phi=gv^2\theta^{45}$, with $v$ the Fermi scale. The
new ingredients in this expression with respect to the standard
result, is the mixing between the field strengths $W_{\mu\nu}^{3}$
and $B_{\mu\nu}$ induced by the $\mathcal{O}_{WB}$ operator, as
well as the presence of a quadratic term in $W_{\mu}^{3}$
generated by the $\mathcal{O}_{\phi}^{(3)}$ operator. As we will
see below, the $W_{\mu\nu}^{3} B^{\mu\nu}$ mixing given by the
$\mathcal{O}_{WB}$ operator contribute to the $S$ parameter at the
tree level, as it involves derivatives. Also, it is important to
notice that while $\mathcal{O}_{\phi}^{(1)}$ affects with the same
intensity both the $W$ and $Z$ masses (see first and second terms
in (\ref{lwb})), the $\mathcal{O}_{\phi}^{(3)}$ operator modifies
only the $Z$--mass, as it is evident from the term proportional to
$W^3_\mu W^3_\nu$. This asymmetric contribution to the gauge field
masses is the responsible for deviations from the SM value of the
$\rho$--parameter $\rho=\alpha T$\cite{EL,DHT}. This contributions
is associated with a violation of the custodial $SU(2)$ symmetry
\cite{Custodial}, which as it is well known, guarantees the tree
level value $\rho=1$ in the SM. The diagonalization of the
resultant kinetic energy sector and its impact on the SM
parameters have been studied in the literature in the more general
context of electroweak effective Lagrangians \cite{EL,Wudka2}. The
tree level contribution of the $\mathcal{O}_{WB}$ and
$\mathcal{O}_{\phi}^{(3)}$ operators to the $S$ and $T$ parameters
has already been studied in this more general context by Hagiwara
\textit{et al.} \cite{H}. For our purposes, it is convenient to
follow the approach introduced by these authors. The oblique
parameters characterize the influence of physics beyond the Fermi
scale. They are given as linear combinations of the transverse components of the gauge--boson
vacuum polarizations
\begin{equation}
\Pi^{\mu \nu}_{ij}(p)=\Pi_{ij}(p^2)g^{\mu\nu}+(p^\mu p^\nu \ {\rm terms}),
\end{equation}
where $ij$ stands for $aa$, $YY$, and $3Y$, where $(a, Y)$ are $SU(2)\times U_Y(1)$ indices.
In particular, the $S$
and $T$ parameters are defined by\cite{DKS}
\begin{eqnarray}
\alpha S&=&\frac{2e^2}{m^2_Z}[\Pi_{3Y}(0)-\Pi_{3Y}(m^2_Z)], \\
\alpha T&=&\frac{2e^2}{m^2_W}{\rm Re}[\Pi_{11}(0)-\Pi_{33}(0)].
\end{eqnarray}
It is not difficult to see from (\ref{lwb}), that
$\mathcal{O}_{WB}$ contributes to $S$ but not to $T$, whereas
$\mathcal{O}^{(3)}_\phi$ contributes to $T$ but not to $S$. It is
also evident from (\ref{lwb}) that $\mathcal{O}^{(1)}_\phi$
does not contribute to these parameters. Considering only the new
physics contribution, one finds,
\begin{eqnarray}
S_{NP}&=&2c_W\sqrt{\frac{\pi}{3\alpha}}(v^2\theta^{45}), \\
T_{NP}&=&-\frac{1}{s_W}\sqrt{\frac{\pi}{\alpha}}(v^2\theta^{45}),
\end{eqnarray}
where $s_W$ and $c_W$ stand for $sin\theta_W$ and $cos\theta_W$,
respectively. On the other hand, the current experimental data
\cite{PDG} give the next values of $S$ and $T$ which can be
induced by new physics effects
\begin{eqnarray}
S^{Exp}_{NP}=-0.13\pm0.10(-0.08), \\
T^{Exp}_{NP}=-0.17\pm 0.12(+0.09),
\end{eqnarray}
where the central values assume $m_H=117$ GeV. The change for
$m_H=300$ GeV is shown in parenthesis. Assuming the first value
for the Higgs mass and using the values for the SM parameters
reported in \cite{PDG}, one finds at $95\%$ C.L.
\begin{eqnarray}
-5.63\times 10^{-8}\ GeV^{-2}&<\theta^{45}<&1.91\times
10^{-7} \ GeV^{-2},\\
-3.18\times 10^{-7} \ GeV^{-2}&<\theta^{45}<&1.06\times 10^{-7} \
GeV^{-2},
\end{eqnarray}
which leads to the following bound for the new physics scale
\begin{equation}
\label{scale} -5.63\times 10^{-8}\ GeV^{-2}<\theta^{45}<1.06\times
10^{-7} \ GeV^{-2}.
\end{equation}
To conclude this part, it is worth comparing this bound with those
obtained in the literature for the noncommutativity scale of four
dimensional theories. To this respect, bounds of order of one TeV
have been obtained from collider physics \cite{BCP}. On the other
hand, more stringent bounds of order $(10\ TeV)^{-2}$ \cite{Carroll}
or higher \cite{Carlson} have been derived from low--energy tests of
Lorentz violation. However, as it has been recently argued
\cite{Calmet}, that these bounds are extremely model dependent and
should be taken with some care.

We are now in position to discuss some phenomenological implications of these
new interactions. Motivated by the fact that new physics effects would be more
evident in those processes which are forbidden or strongly suppressed in the
SM, we will consider the rare Higgs boson decay into two photons, which is an
one--loop prediction of the model and thus is naturally suppressed \cite{HSM}.
Due to its phenomenological importance, this decay has been the subject of
permanent interest in the literature. Apart from providing a good signature
for the Higgs boson search at hadron colliders with mass in the intermediate
range $120\ \mathrm{Gev}<m_{H}<2m_{Z}$ \cite{DHC}, the decay width of this
process is also of great interest because it determines the cross section for
Higgs production in $\gamma\gamma$ collisions \cite{PPP}. Due to the fact that
the $H\gamma\gamma$ coupling is generated by loop effects of charged
particles, its sensitivity to new heavy charged particles has been studied in
many well motivated extensions of the SM, as the two Higgs doublet model
(THDM) \cite{HHG}, the minimal supersymmetric standard model (MSSM)
\cite{SUSY}, the left--right symmetric models (LRM) \cite{LR}, and the
Littlest Higgs model (LHM) \cite{Han}. Many of its properties have also been
studied in a model--independent manner using the effective Lagrangian
framework \cite{TH,HEL}. In our model, the $H\gamma\gamma$ vertex (as well as
$H\gamma Z$ one) is induced at the tree level by the set of operators given in
eqs.(\ref{o1}-\ref{o3}). The corresponding Lagrangian can be written as
follows:
\begin{equation}
\mathcal{L}_{H\gamma\gamma}=\frac{\alpha_{W}}{4}m_{W}\theta^{45}HF_{\mu\nu
}F^{\mu\nu},
\end{equation}
where $\alpha_{W}=4s_{W}^{2}-c_{W}^{2}-2s_{2W}/\sqrt{3}$. The total decay
width $\Gamma_{NC}$ can be conveniently written in terms of the SM width
$\Gamma_{SM}$, as follows:
\begin{equation}
\Gamma_{NC}(H\rightarrow\gamma\gamma)=\Gamma_{SM}(H\rightarrow\gamma
\gamma)\Big|1+\frac{\mathcal{A}_{NC}}{\mathcal{A}_{SM}}\Big|^{2},
\end{equation}
where $\mathcal{A}_{NC}=\alpha_{W}m_{W}^{2}\theta^{45}$, whereas
$\mathcal{A}_{SM}$ represents the charged fermion and $W$ boson loop
contributions, which is given by
\begin{equation}
\mathcal{A}_{SM}=\frac{\alpha^{3/2}}{\sqrt{4\pi}s_{W}}\Big[\sum_{f}N_{Cf}%
Q_{f}^{2}F_{f}+F_{W}\Big].
\end{equation}
In this expression, $f$ stands for quarks or leptons, $N_{Cf}$ is the color
index, and $Q_{f}$ is the electric charge of the fermion in units of the
charge of the positron. In addition, $F_{f}$ and $F_{W}$ are the fermion and
$W$ boson loop amplitudes, respectively, which can be found in ref.
\cite{HHG}.
\begin{figure}[ptb]
\centering \includegraphics[width=3in]{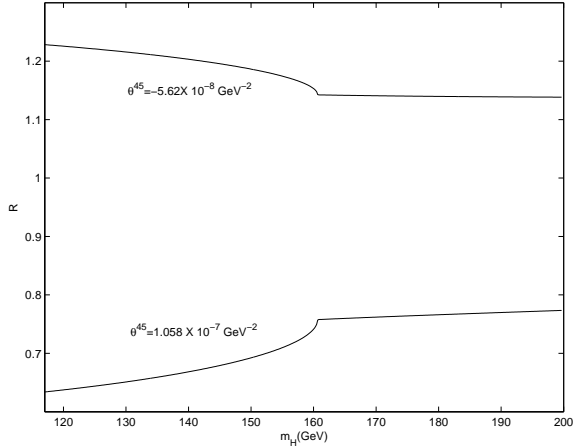}
\caption{The
$R=\Gamma _{NP}(H\to\gamma\gamma)/\Gamma_{SM}(H\to\gamma\gamma)$
ratio as function of the
Higgs mass. }%
\label{FIG}%
\end{figure}
Though the bound for $\theta^{45}$ was estimated for a value
$m_H=117$ GeV, for illustration purposes we will present results
that contemplate larger values of the Higgs mass. In
Fig.\ref{FIG}, the variation of the normalized decay width
$R=\Gamma_{NC}/\Gamma_{SM}$ is displayed as function of the Higgs
mass. From this figure, it can be appreciated that $R$ is
sensitive to the sign of the $\theta^{45}$ parameter. A
constructives or destructives effect corresponds to
$\theta^{45}<0(\theta^{45}>0)$, which can increase or decrease the
standard model prediction $\Gamma_{SM}$ up to by $27\%$ and
$35\%$, respectively, for $m_{H}$ in the range $120-200$ GeV.

It is interesting to compare this result with those obtained from
other models. In general, theories beyond the SM require more
complicated Higgs sectors, \textit{i.e.} they incorporate new
neutral and charged Higgs bosons. However, in most cases, it is
always possible to identify in an appropriate limit a SM--like Higgs
boson, that is a CP--even neutral scalar whose couplings to pairs of
$W$ and $Z$ bosons coincide with those given in the minimal SM.
Furthermore, new contributions coming from new charged scalar,
fermion, and vector particles are expected. We briefly review the
results for a SM--like Higgs boson decaying into two photons, for
the models mentioned previously. In the THDM, the only new charged
particle is the $H^{\pm}$ Higgs boson, but its loop contribution to
this decay is very small compared with the dominant $W$
contribution. Thus, in this model, the $\gamma\gamma$ decay width is
essentially the same as for the SM \cite{HHG}. In the case of the
MSSM, this decay gets new contributions from superpartner loops. The
$\gamma\gamma$ width tends to be lower than the one of the SM due to
cancelations between the $W$ loop and the supersymmetric chargino
loops. In this case, if the charginos are heavy, the decay width can
be quite near to the SM width \cite{SUSY}. Further, it was found in
ref. \cite{LR} that the contribution of a new $W$ boson, like the
one predicted by LRM, is quite suppressed, because the corresponding
loop amplitude is related to the SM $W$ boson amplitude as
$F_{W_{R}}=(m_{W_{L}}/m_{W_{R}})^{2}F_{W_{L}}$. Taking into account
the existing $W_{R}$--mass bounds \cite{PDG}, the $\gamma\gamma$
width can be enhanced up to $5\%$ at best. As to the prediction of
the Littlest Higgs model is concerned, in \cite{Han} it was found
that the $\gamma\gamma$ width is reduced by $5-7\%$ compared to the
SM value. From these results, we can see that the impact of
noncommutative extra dimensions on the $H\to\gamma\gamma$ decay may
be significantly more important than that predicted by some of the
most popular SM extensions.

%%%%%%%%%%%%%%%%%%%%%%%%%%%%%%%%%%%%%%%%%%%%%%%%%%%%%%%%%%%%%%%%%%%%%%%%%%%%%%

\section{Conclusions}

In this work we explore the consequences of noncommutativity in a
6-dimensional model, by means of the Seiberg-Witten map. We consider the
$SU(3)$ gauge Higgs unification model of the electroweak interactions of
\cite{antoniadis1} and \cite{Wulzer}, compactified to 4D on an orbifold
$T^{2}/Z_{N}$ for $N=3,4,6$. We
analyze noncommutativity among all the 6-dimensional coordinates. As a
consequence of the orbifold symmetries, it turns out that there are no
corrections to the model due to noncommutativity among the 4D coordinates and
the two-extra dimensions. We find that the corrections we obtain corresponding
to noncommutativity among the 4D coordinates,
differ from the ones of noncommutative models calculated directly in 4D, also
by means of the Seiberg-Witten map \cite{W3}. On the other side, the corrections
corresponding to noncommutativity of the extra dimensions have interesting
phenomenological consequences, as we emphasize below.

As well as in the commutative
model, the spontaneous symmetry breaking should arise dynamically, from first
order quantum corrections. This step can be done in the noncommutative theory,
as far as the expected Higgs mass is much less than the noncommutativity scale.
Thus it would be interesting to include matter and
to study the corresponding noncommutative corrections, which could be done
following \cite{W2}, progress in this direction will be reported elsewhere.

As mentioned in the introduction, the model we are considering
here has a too high value for the weak angle. However, as noted in
\cite{Wulzer}, there are various ways to solve this problem, in particular
by an extension by a $U(1)$ factor. Furthermore, the noncommutative
Seiberg-Witten map of the corresponding gauge field will not mix with the
already present noncommutative corrections.
Thus we can expect that in a noncommutative version of this extended
model, the kind of corrections presented here will still be
present, in particular those corresponding to noncommutativity
between extra dimensions. Under this working assumption, we have
studied the corrections due to noncommutativity of the extra dimensions
by means of the effective lagrangian
techniques. First, by the observation that these terms are quite
sensitive to the $S$ and $T$ oblique parameters, we could obtain
the bound to $\theta^{45}$ given by (\ref{scale}). In four dimensional
noncommutative models, there are bounds obtained \textit{e.g.}
from low--energy tests of Lorentz violation, which are extremely
model dependent \cite{Calmet}. We think that, in the framework of
our working assumption, our bound has a less speculative nature,
as it was obtained directly from the experimental constraints on
the oblique parameters, without additional assumptions. With this
bound established, we have looked at the impact of our corrections
on the rare Higgs decay into two photons. It turns out that the
effect depends on the signature of the noncommutativity parameter,
increasing or decreasing respectively the value of the SM decay
width $\Gamma_{SM}$, with a net effect which could be
significantly more important than that predicted by some of the
most popular SM extensions \cite{PDG,SUSY,LR,Han}.

Finally, from the results of the particular noncommutative model we started
with, which could be interesting on its own, we can conclude that
noncommutativity in higher dimensional models can have interesting
consequences and phenomenological effects beyond those of four dimensional
noncommutative theories. The study of more realistic models, including matter
fields, is in progress.
%%%%%%%%%%%%%%%%%%%%%%%%%%%%%%%%%%%%%%%%%%%%%%%%%%%%%%%%%%%%%%%%%%%%%%%%%%%%%%

\begin{acknowledgments}
We thank H. Garc\'{\i}a-Compe\'{a}n for discussions. This work has
been supported by CONACYT grant 47641 and Projects by PROMEP, UG and
VIEP-BUAP 13/I/EXC/05.
\end{acknowledgments}

%%%%%%%%%%%%%%%%%%%%%%%%%%%%%%%%%%%%%%%%%%%%%%%%%%%%%%%%%%%%%%%%%%%%%%%%%%%%%%

\end{document}